\documentclass[superscriptaddress, 12pt]{revtex4}

\usepackage{graphicx}
\usepackage{times}
\usepackage{bm}
\usepackage{amsmath}
\usepackage{natbib}

\usepackage[colorlinks = true,
            linkcolor = blue,
            urlcolor  = blue,
            citecolor = blue,
            anchorcolor = blue]{hyperref}

\newcommand{\changeurlcolor}[1]{\hypersetup{urlcolor=#1}}

\begin{document}

\title{u-Channel Color Transparency Observables}

\author{G.M. Huber}
\email[]{huberg@uregina.ca}
\affiliation{University of Regina, Regina SK  S4S~0A2 Canada}
\author{W.B. Li}
\affiliation{Center for Frontiers in Nuclear Science, Stony Brook University, Stony Brook, NY 11794, USA}
\affiliation{Department of Physics and Astronomy, Stony Brook University, Stony Brook, NY 11794, USA}
\author{W. Cosyn}
\affiliation{Department of Physics, Florida International University,  Miami, Florida 3199, USA}
\affiliation{Department of Physics and Astronomy, Ghent University, B9000 Gent, Belgium}
\author{B. Pire}
\affiliation{CPHT, CNRS, \'Ecole polytechnique, I.P. Paris, F91128 Palaiseau, France}

\begin{abstract}
We propose to study the onset of color transparency in hard exclusive reactions in the backward regime. Guided by the encouraging JLab results on backward $\pi$ and $\omega$ electroproduction data at moderate $Q^2$, which may be interpreted as the signal of an early scaling regime where the scattering amplitude factorizes in a hard coefficient function convoluted with nucleon to meson transition distribution amplitudes, we show that the study of these channels on nuclear targets opens a new opportunity to test the appearance of nuclear color transparency for a fast-moving nucleon.
\end{abstract}

\maketitle

\section{Introduction}

\begin{figure}
\begin{center}
\includegraphics[width=0.4\textwidth]{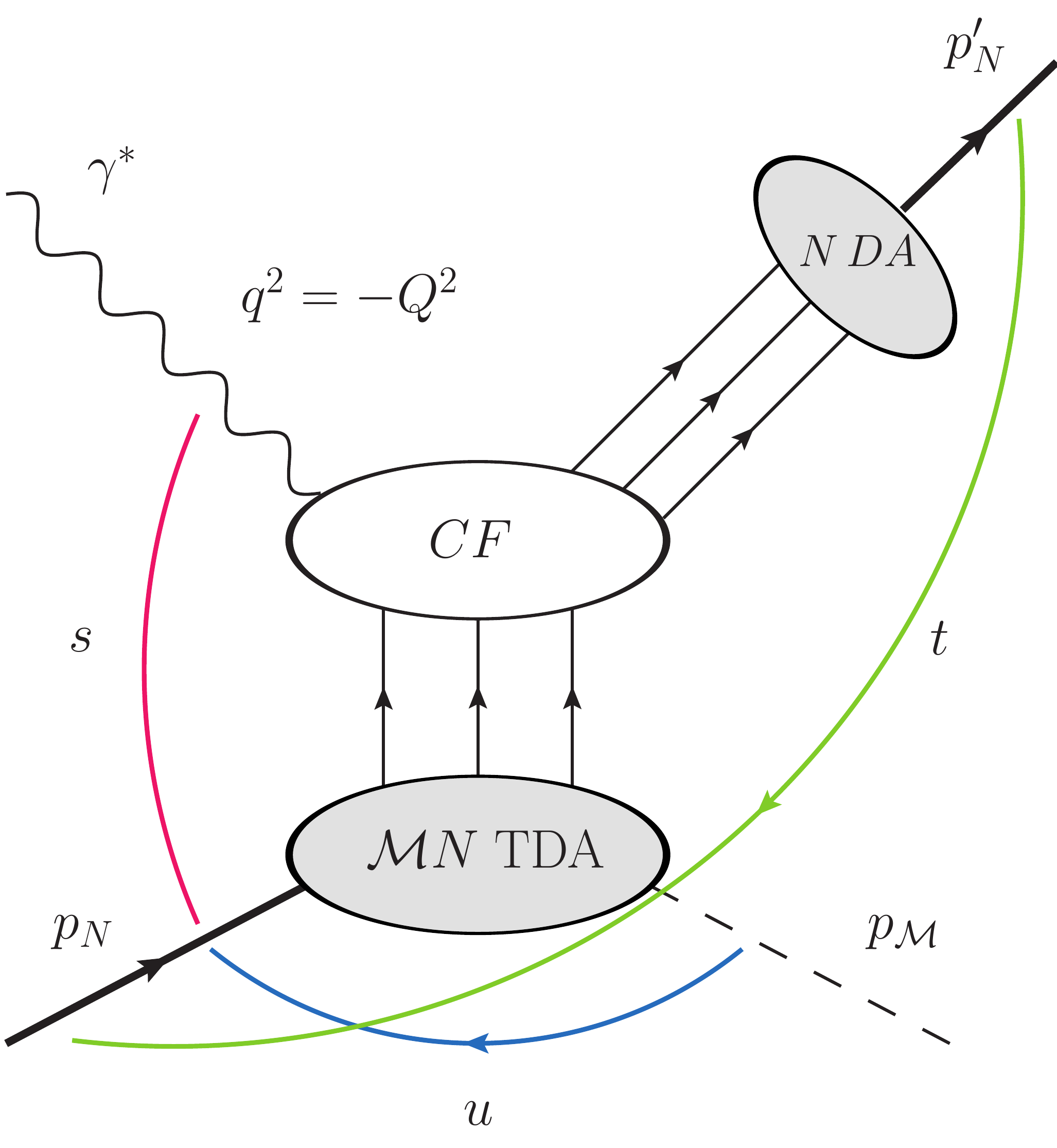}
\end{center}
     \caption{Kinematical quantities and the collinear factorization mechanism  for
      $\gamma^* N \to N {\cal M}$  in the  near-backward  kinematical regime (large $Q^2$, $W$; fixed $x_{B}$; $|u| \sim 0$). The lower blob, denoted
  ${\cal M}N$ TDA, depicts the nucleon-to-meson~$\cal M$ transition
     distribution amplitude; the $N$ DA blob depicts the nucleon distribution amplitude;
      $CF$  denotes the hard subprocess amplitude (coefficient function).}
\label{Fig_Kinematics_TDAs}
\end{figure}

While being a fundamental prediction of QCD \cite{Mueller:1982bq, Brodsky:1982kg}, the phenomenon of color transparency has been for many decades a domain of controversial interpretations of experimental data (for a review, see e.g. \cite{Jain:1995dd}). Together with scaling laws and polarization tests, the increase of nuclear transparency (NT) ratio with the relevant hard scale (denoted as $Q^2$) is believed to constitute an important signal of the onset of a collinear QCD factorization regime where hadrons transverse sizes shrink proportionally to $\frac{1}{Q}$, thus drastically diminishing final-state interaction cross-sections.

Near-forward exclusive photon or meson electroproduction processes have been the subject of intense theoretical and experimental studies \cite{Diehl:2003, Kumericki:2016ehc}. Most of the available data are now interpreted in terms of a collinear QCD factorized amplitude, where generalized parton distributions (GPDs) are the relevant hadronic matrix elements. The study of nuclear transparency for meson electroproduction \cite{Clasie:2007aa,CLAS:2012tlh} indeed revealed a growth of the NT ratio indicative of an early on-set of the scaling regime. This may however look contradictory to the non-dominance of the leading twist pion production amplitude revealed by the small value  of the polarization ratio $\sigma_L / \sigma_T$ for this reaction \cite{JeffersonLabHallA:2016wye, JeffersonLabHallA:2020dhq}. Alternative models have been proposed \cite{Kaskulov:2010} to explain this fact.

Exclusive electroproduction processes in the complementary near backward region (see Fig.\ref{Fig_Kinematics_TDAs}), where $- u = -(p_M - p_N)^2 \ll Q^2$ is near its minimal value~\cite{Gayoso:2021rzj}, should also be described at large $Q^2$ in a collinear QCD factorization scheme \cite{Frankfurt:1999fp, Pire:2004ie, Pire:2005ax}, where nucleon to meson transition distribution amplitudes (TDAs) replace the GPDs as the relevant hadronic matrix elements \cite{Pire:2021hbl}. Indeed, the first experimental studies \cite{Park:2017irz, Li:2019xyp, Diehl:2020uja} of this new domain at rather moderate values of $Q^2$ point toward an early onset of the scaling regime.

\section{The proposed measurement}
\subsection{Previous backward-angle data from Jefferson Lab}

The first data providing qualitative support for the TDA picture are from the JLab 6~GeV physics program \cite{Park:2017irz, Diehl:2020uja, Li:2019xyp}.

\begin{figure} 
\centering
  \includegraphics[width=0.6\textwidth]{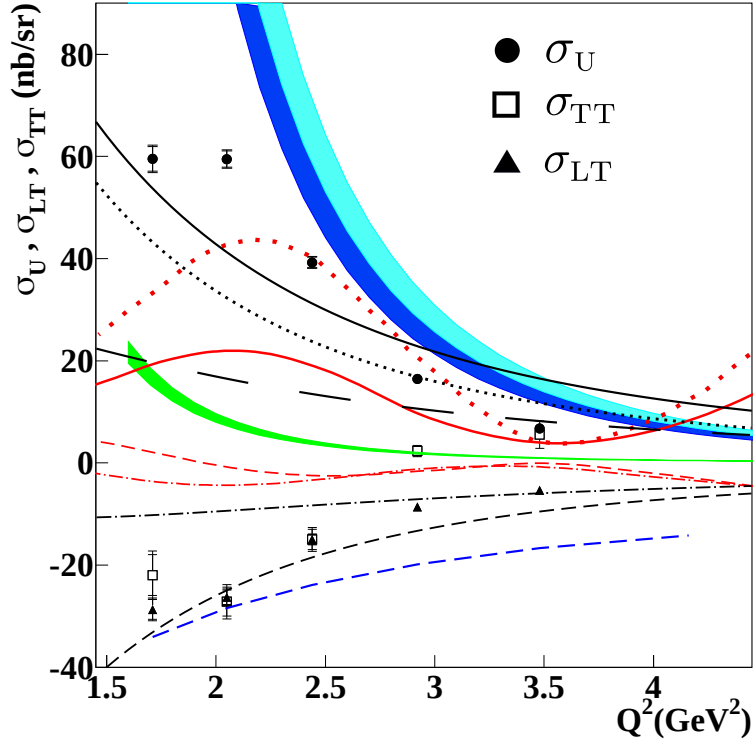}
  \caption{The structure functions $\sigma_U = \sigma_T+\varepsilon \sigma_L$ (solid dot), $\sigma_{TT}$ (square) and $\sigma_{LT}$ (triangle) as a function of $Q^2$. The inner error bars are statistical and the outer error bars are the combined systematic and statistical uncertainties in quadrature.  The bands refer to model calculations of $\sigma_U$ in the TDA description, green band: BLW NNLO, dark blue band: COZ, and light blue band: KS (see \cite{Park:2017irz} and refs. therein for the meaning of these models). The lower blue short-dashed line represents an educated guess to fit the higher twist cross-sections $\sigma_{LT}$ and $\sigma_{TT}$ in the TDA picture. The red curves are the ``Regge'' predictions (by JML18) of \cite{guidal97, Laget:2019} for solid: $\sigma_U$, dashed curve: $\sigma_{LT}$, dot-dashed: $\sigma_{TT}$. An updated $\sigma_U$ calculation from JML18 model ~\cite{Laget:2021qwq} are shown in red dotted curve. Regge calculations which consider parton contributions (see \cite{workshop2021}) to $\sigma_{U}$, $\sigma_{T}$, $\sigma_{L}$, $\sigma_{TT}$ and $\sigma_{LT}$ are shown in black solid, black dotted, black  long-dashed, black dot-dashed and black short-dashed, respectively. This plot was recreated from the data and model predictions published in Ref.~\cite{Park:2017irz, workshop2021}.   
    }
  \label{fig:CLAS_cross_section_u_channel} 
\end{figure}

\begin{figure} 
\centering
\hspace{0mm}
  \includegraphics[width=0.6\textwidth]{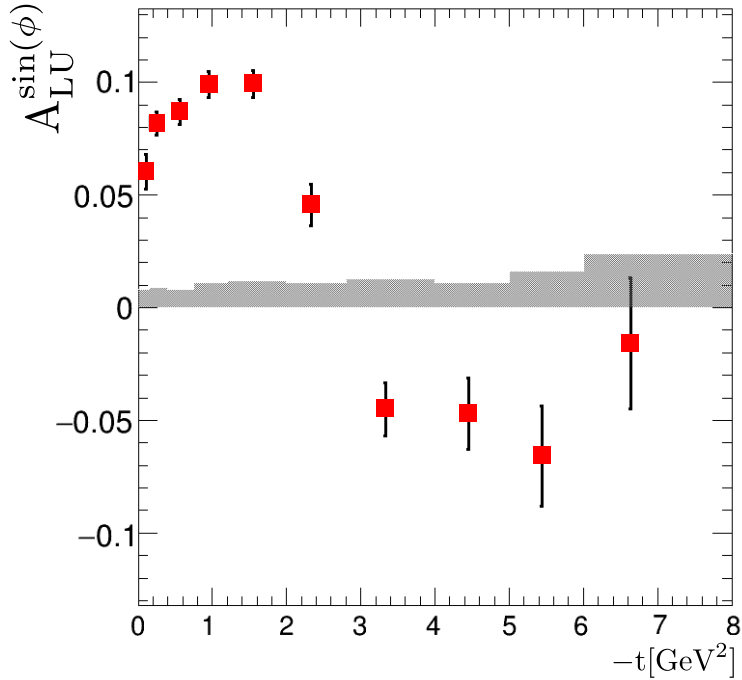}
\caption{$A_{LU}^{\sin\phi}$ as function of $-t$ measured with CLAS at $W >$ 2 GeV, $Q^{2} >$ 1 GeV$^2$. The maximal accessible value of $-t$ is $\approx 8.8$ GeV$^{2}$. The shaded area represents the systematic uncertainty. This plot was recreated from the data and model predictions published in Ref.~\cite{Diehl:2020uja}.}
\label{fig:CLAS_t_ALU_PRL} 
\end{figure}

\begin{figure}
\centering
\includegraphics[width=0.9\textwidth]{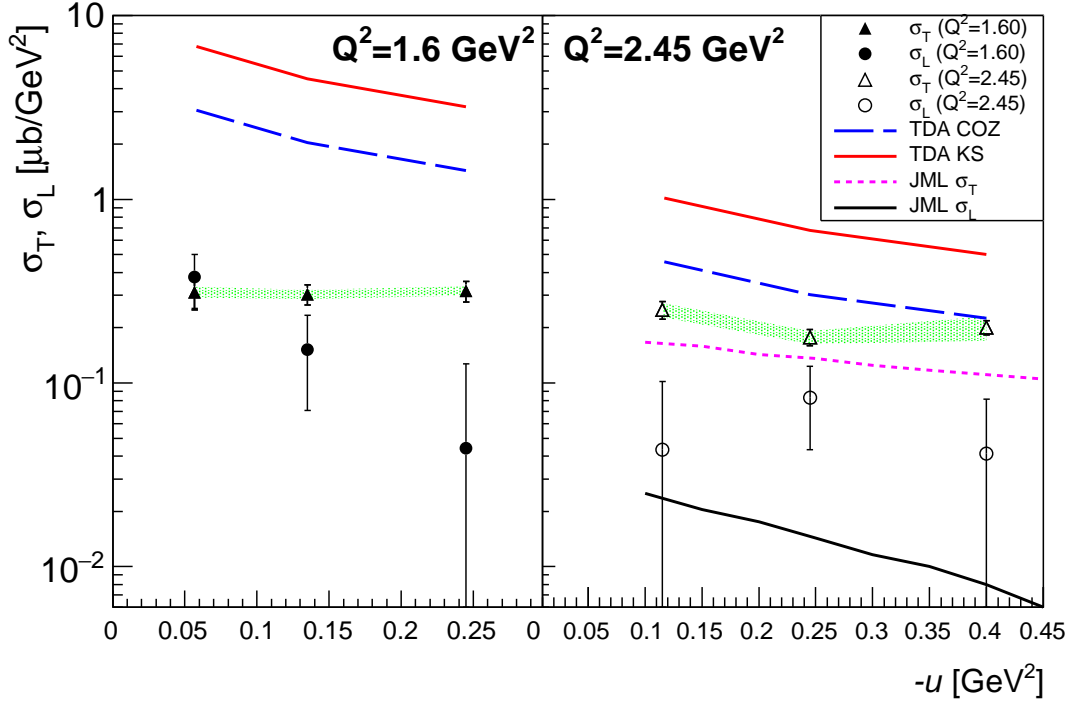}
\caption{$\sigma_{T}$ (triangles), $\sigma_{L}$ (squares) as a function
  of $-u$, at $Q^2=1.6$ GeV$^2$ (left), 2.45 GeV$^2$ (right). 
   For the lowest $-u$ bin, $u^{\prime} = u- u_{\rm min}\approx0$. 
   TDA predictions for $\sigma_{T}$: COZ (blue dashed
   lines), KS (red solid lines). The predictions were
   calculated at the specific $\overline{Q^2}$, $\overline{W}$ values of each
   $u$ bin, and the predictions at three $u$ bins joined by straight lines for
   visualization purpose.  Green bands indicate correlated systematic
   uncertainties for $\sigma_{T}$, the uncertainties for $\sigma_{ L}$ are similar. 
   This plot was recreated from the data and model predictions published in Ref.~\cite{Li:2019xyp}.}
   \label{fig:sigt}
\end{figure}

Hard exclusive $\pi^{+}$ production ($e p \rightarrow e' n \pi^+$) from a polarized electron beam interacting with an unpolarized hydrogen target was studied with the CLAS detector in the backward angle kinematic regime by Park et al. \cite{Park:2017irz}. Fig.~\ref{fig:CLAS_cross_section_u_channel} shows the $Q^2$-dependence of $\sigma_U=\sigma_T+\varepsilon\sigma_L$, $\sigma_{LT}$ and $\sigma_{TT}$, obtained at the average kinematics of $W=2.2$ GeV and $-u=0.5$ GeV$^2$. All three cross-sections have a strong $Q^2$-dependence. The TDA formalism predicts that the transverse amplitude dominates 
at large $Q^2$. 
With only this set of data at fixed beam energy, the CLAS detector cannot experimentally separate $\sigma_T$ and $\sigma_L$. After examining the angular dependence of $\sigma_U$, the results show $\sigma_{TT}$ and $\sigma_{LT}$ roughly equal in magnitude and with a
similar $Q^2$-dependence. Their significant sizes (about 50\% of $\sigma_U$) imply an important contribution of the transverse amplitude in the cross-section.  Furthermore, above $Q^2=2.5$ GeV$^2$, the trend of $\sigma_U$ is qualitatively consistent with the TDA calculation, yielding the characteristic $1/Q^8$ dependence expected when the backward collinear factorization scheme is approached.

The beam spin asymmetry moment, $A_{LU}^{\sin\phi}$, was also extracted using the CLAS detector \cite{Diehl:2020uja}.
$A_{LU}^{\sin\phi}$ is proportional to the polarized structure function $\sigma_{LT^\prime}$,
\begin{equation}\label{eq:ALU}
\begin{aligned}
\hspace{6em} 	A_{LU}^{\sin\phi} = \frac{\sqrt{2 \varepsilon (1 - \varepsilon)}~\sigma_{LT^{\prime}}}{\sigma_{T} + \varepsilon \sigma_{L}},
\end{aligned}
\end{equation}
where the structure functions $\sigma_{L}$ and $\sigma_{T}$ correspond to longitudinal and transverse virtual photons, and $\varepsilon$ describes the ratio of their fluxes.
Due to the large acceptance of CLAS, it was possible to map out the full kinematic region in $-t$ from very forward kinematics ($-t/Q^{2} \ll 1$) where a description based on Generalized Parton Distributions (GPD) can be applied, up to very backward kinematics ($-u/Q^{2} \ll 1$, $-t$ large),
where a TDA-based description is expected to be valid. The results in Fig. \ref{fig:CLAS_t_ALU_PRL} indicate a clear transition from positive $A_{LU}^{\sin\phi}$ in the forward regime to rather small negative values in the backward regime, with a $Q^2$ dependence qualitatively consistent with the TDA picture. The sign change between the forward and backward kinematic regimes is independent of $Q^2$ and $x_B$ within the kinematics accessible with CLAS.

Backward-angle exclusive $\omega$ electroproduction ($e p \rightarrow e^{\prime} p \omega$), was
studied in Hall C by Li et al. \cite{Li:2019xyp}. The scattered electron and forward-going
proton were detected in the HMS+SOS spectrometers, and the
low momentum rearward-going $\omega$ was reconstructed using the missing mass
reconstruction technique. Since this method does
not require the detection of the produced meson, it allows the analysis to
extend the experimental kinematics coverage to a region that is inaccessible through the standard direct detection method.
The extracted $\sigma_{\rm L}$ and $\sigma_{\rm T}$ as a function of $-u$ at
$Q^2=1.6$ and 2.45~GeV$^2$ are shown in Fig.~\ref{fig:sigt}.  The two sets of
TDA predictions \cite{Pire:2015kxa} for $\sigma_{\rm T}$ each assume different nucleon DAs \cite{Chernyak:1987nv, King:1986wi} as
input.  From the general trend, the TDA model offers a good description of the
falling $\sigma_{\rm T}$ as a function of $-u$ at both $Q^2$ settings, similar to the backward-angle $\pi^+$ data from \cite{Park:2017irz}.
Together, the data sets are suggestive of early TDA scaling.
The behavior of $\sigma_{\rm L}$ differs greatly at the two $Q^2$
settings. At $Q^2=$1.6~GeV$^2$, $\sigma_{\rm L}$ falls almost exponentially as
a function of $-u$; at $Q^2=$2.45~GeV$^2$, $\sigma_{\rm L}$ is constant
near zero (within one standard deviation). Note that the TDA model predicts a small -- higher twist -- $\sigma_{\rm L}$ contribution, which falls faster with $Q^2$ than the leading twist $\sigma_{\rm T}$ contribution.

\subsection{Jefferson Lab proposal E12-20-007}

\begin{figure}
\centering
\includegraphics[width=0.9\textwidth]{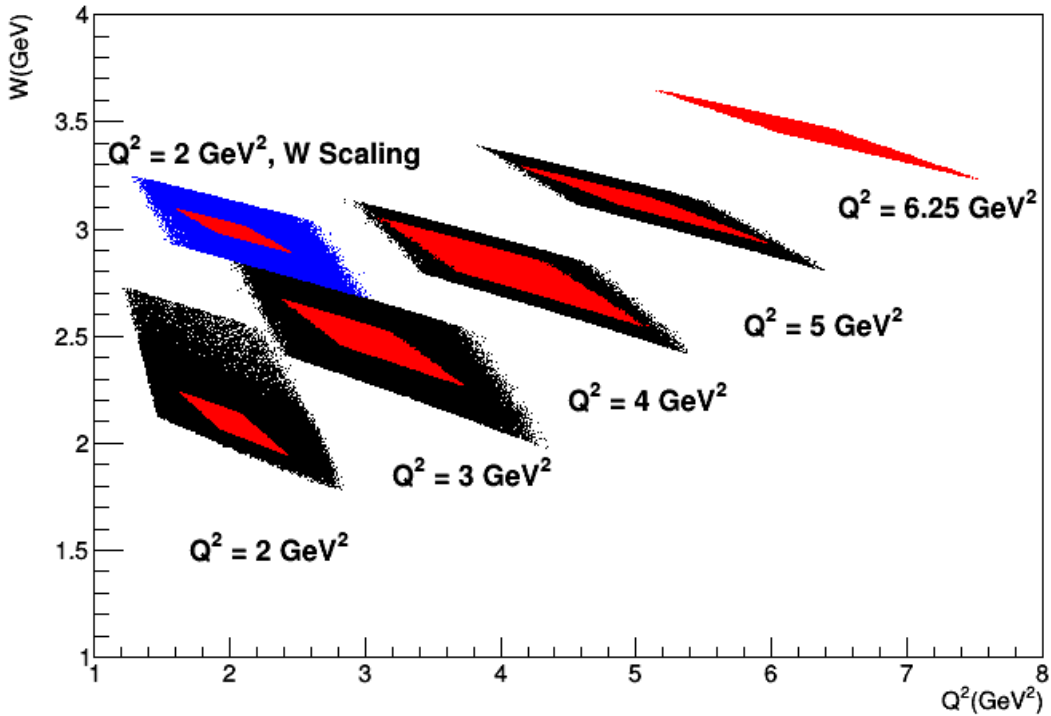}
\caption{$W$ vs $Q^2$ diamonds for the $Q^2=2.0$, $3.0$, $4.0$, $5.0$ and $6.25$ GeV$^2$ settings of E12-20-007. The black diamonds are for the higher $\epsilon$ settings and the red diamonds are for the lower $\epsilon$ settings. The diamonds for the $W$ scaling setting are shown separately (in blue and red). Note that there is only one $\epsilon$ setting for $Q^2=6.25$ GeV$^2$. The overlap between the black and red diamond is critical for the L/T separation at each setting. The boundary of the low $\epsilon$ (red) data coverage will become a cut for the high $\epsilon$ data.}
\label{fig:diamond}
\end{figure}

\begin{figure}
    \centering
    \includegraphics[width=0.495\textwidth, height=0.495\textwidth]{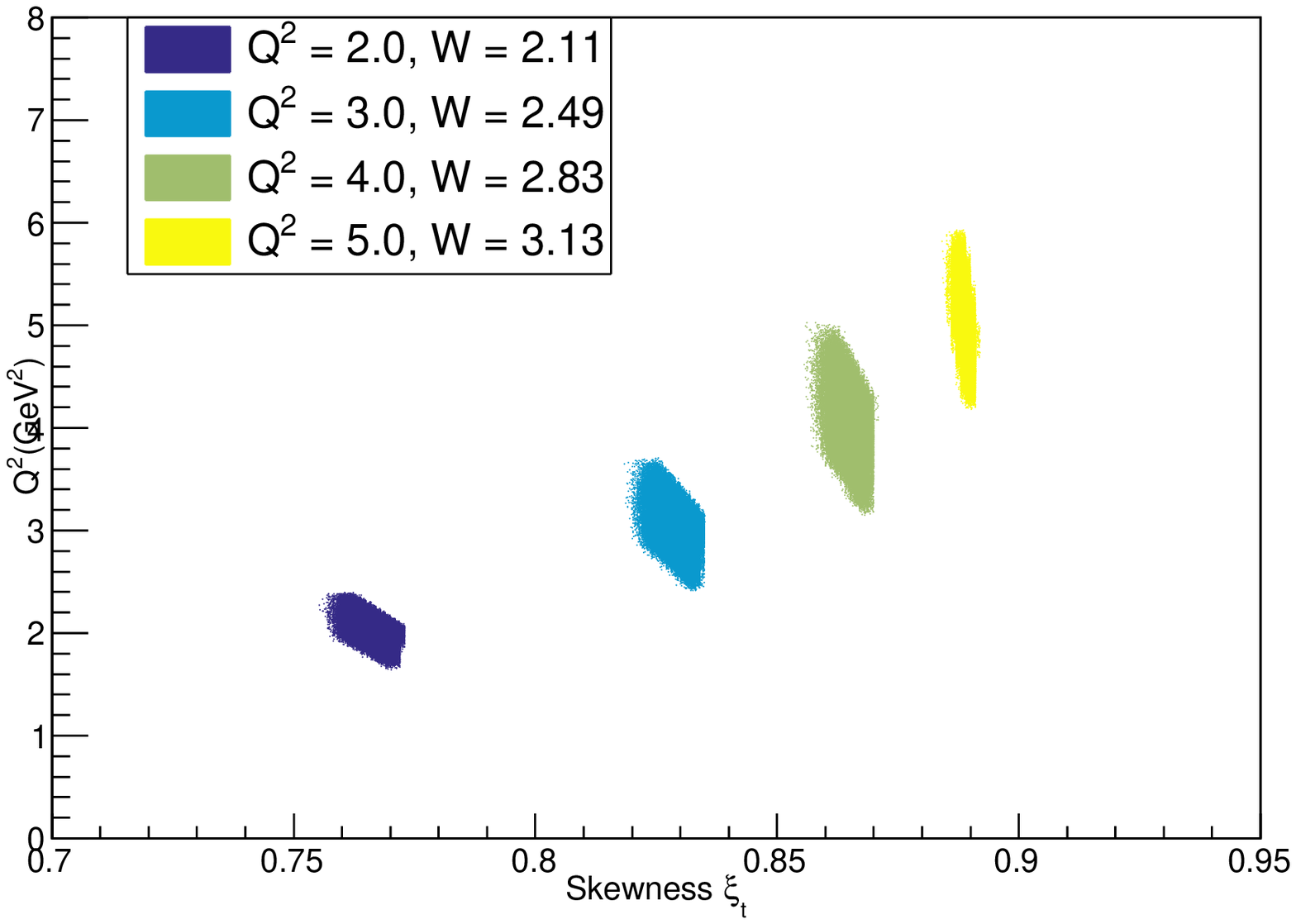}
    \includegraphics[width=0.495\textwidth, height=0.495\textwidth]{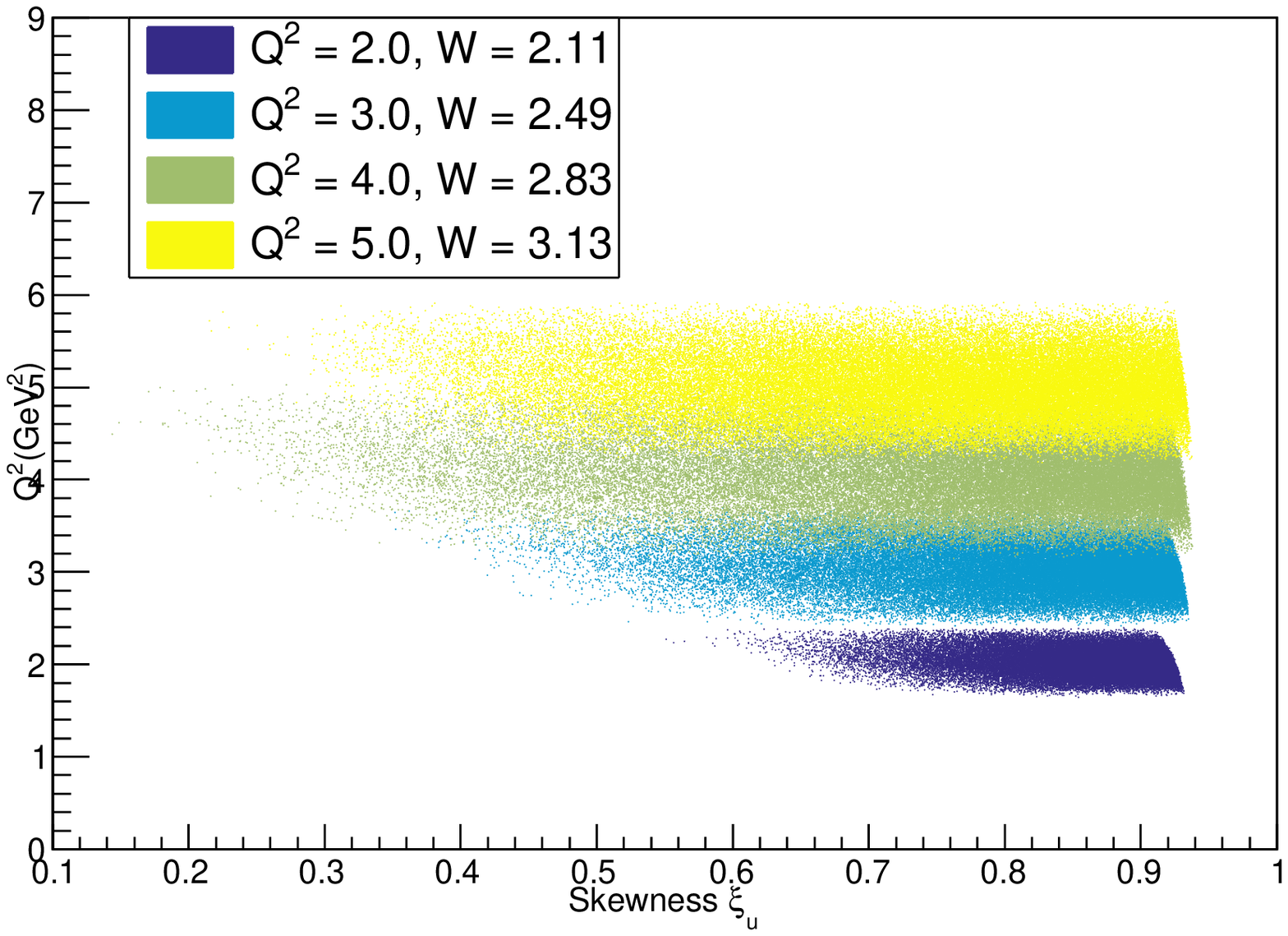}
    \caption{Left: forward skewness $\xi_t$, Right: backward skewness $\xi_u$ coverage of the planned E12-20-007 measurements for $Q^2=2.0$ to 5.5~GeV$^2$ (see text for definitions.)  Clearly the experiment probes a wide kinematic range, which will be helpful for distinguishing the roles of the TDA and Regge reaction mechanisms in the soft-hard transition range.}
    \label{fig:skewness}
\end{figure}

The first dedicated experiment for exclusive $\pi^0$ production in backward kinematics ($e p \rightarrow e^{\prime} p \pi^0$),
was proposed by Li, et al. in \cite{Li:2020nsk}.
Here, the produced $\pi^0$ is emitted 180 degrees opposite to the virtual-photon momentum (at large momentum transfer), and is reconstructed via the missing mass technique, just as in \cite{Li:2019xyp}. This study aims to apply the Rosenbluth separation technique to provide model-independent (L/T) differential cross-sections at the never explored $u$-channel kinematics region near  ($-t=-t_{max}$, $-u=-u_{min}$). 

The kinematic coverage of the experiment is shown in Fig.~\ref{fig:diamond}.  The L/T-separated cross-sections are planned at $Q^2=2.0$, $3.0$, $4.0$ and $5.0$~GeV$^2$. These measurements will provide the $-u$ dependence for $\sigma_{\rm L}$ and $\sigma_{\rm T}$ at nearly constant $Q^2$ and $W$, in addition to the behavior of $\sigma_{\rm L}$/$\sigma_{\rm T}$ ratio as function of $Q^2$. The $Q^2=6.25$~GeV$^2$ setting is chosen to test the $Q^2$ scaling nature of the unseparated cross-section, but only one $\epsilon$ setting is available due to limitations on the accessible spectrometer angles. 
These measurements are intended to provide a direct test of two predictions from the TDA model \cite{Lansberg:2011aa}: $\sigma_T \propto 1/Q^8$ and $\sigma_T \gg \sigma_L$ in $u$-channel kinematics.
 The magnitude and $u$-dependence of the separated cross-sections also provide direct connections to the re-scattering Regge picture \cite{Laget:2021qwq}. The extracted interaction radius (from $u$-dependence) at different $Q^2$ can be used to study the soft-hard transition in the $u$-channel kinematics. 
 
 Relevant to this discussion is the definition of skewness. For forward-angle kinematics, in the regime where the handbag mechanism and GPD description may apply, the skewness is defined in the usual manner,
\begin{equation}
\xi_t=\frac{p_1^+-p_2^+}{p_1^++p_2^+},
\label{eqn:xi_t}
\end{equation}
where $p_1^+$, $p_2^+$ refer to the light-cone plus components of the initial and final proton momenta \cite{Diehl:2003}.  The subscript $t$ has been added to indicate that this skewness definition is typically used for forward-angle kinematics, where $-t\rightarrow -t_{min}$.  In this regime, $\xi_t$ is related to Bjorken-$x$, and is approximated by $\xi_t=x/(2-x)$, up to corrections of order $t/Q^2 < 1$ \cite{Diehl:2003}.  This relation is an accurate estimate of $\xi_t$ to the few percent level for forward-angle electroproduction.  
In backward-angle kinematics, where $-t\rightarrow -t_{max}$ and $-u\rightarrow -u_{min}$, the skewness is defined with respect to $u$-channel momentum transfer in the TDA formalism \cite{Pire:2021hbl},
\begin{equation}
\xi_u=\frac{p_1^+-p_{\pi}^+}{p_1^++p_{\pi}^+}.
\label{eqn:xi_u}
\end{equation}
Figure \ref{fig:skewness} shows the forward $\xi_t$ and backward $\xi_u$ skewness coverage of the approved measurements.  
The ``soft-hard transition'' in $u$-channel meson production is an interesting and unexplored subject.
 The acquisition of these data will be an important step forward in validating the existence of a backward factorization scheme  of the nucleon structure function and establishing its applicable kinematic range. 

\subsection{Nuclear targets}

Since the Jefferson Lab 6 GeV backward-angle data are qualitatively consistent with early factorization in backward kinematics, backward-angle meson production events with a high momentum forward proton may provide an alternate means of probing Color Transparency.

\begin{table}
    \centering
    \begin{tabular}{|p{1.0cm}|p{1.9cm}|p{2.0cm}|p{2.0cm}|p{1.9cm}|p{1.6cm}|}
    \hline
    \multicolumn{6}{|c|}{$A(e,e'p)\pi^0$ kinematics for $E_{beam}=10.6$ GeV, $W=2$ GeV}\\ 
    \hline
  $Q^2$ (GeV$^2$) & $e^{\prime}$ (GeV/c @ deg) & $p$ (GeV/c @ deg) & $\pi^0$ (GeV/c @ deg) & $t$ (GeV$^2$) & $u$ (GeV$^2$) \\ 
  \hline
  3     & 7.3 @ 11.3$^\circ$ & 3.9-3.6 @ 23$^\circ$-30$^\circ$ & 0.2-0.5 @ 202$^\circ$-95$^\circ$ & -5.7 to -5.2 & +0.5 to -0.1\\ \hline
  6     & 5.7 @ 18.1$^\circ$ & 5.6-5.2 @ 19$^\circ$-24$^\circ$ & 0.1-0.5 @ 196$^\circ$-79$^\circ$ & -8.8 to -8.2 & +0.6 to 0.0\\ \hline
  10    & 3.6 @ 29.7$^\circ$ & 7.7-7.3 @ 13$^\circ$-16$^\circ$ & 0.0-0.5 @ 193$^\circ$-61$^\circ$ & -12.8 to -12.1 & +0.6 to -0.1\\ \hline
    \end{tabular}
    \caption{Possible kinematics for a backward angle Color Transparency experiment in Hall C of Jefferson Lab.  Columns 2-4 indicate the approximate momentum and angle of the detected scattered electron and proton, and the undetected $\pi^0$.  Columns 5 and 6 indicate the expected $t$, $u$ ranges covered by the data.}
    \label{tab:CTkin}
\end{table}

We take exclusive $\pi^0$ production as an example reaction, based on the kinematics of E12-20-007, but the technique is in principle extendable also to vector meson production.  In this case, the scattered electron would be detected in the HMS and the high momentum forward-going proton detected in the SHMS, with the meson reconstructed via the missing mass technique.  Table \ref{tab:CTkin} shows an example  of kinematics (at JLab 12 GeV) for the described measurement.  A comprehensive experiment should cover a range of nuclear targets, such as $^{1,2}$H, $^{12}$C, $^{27}$Al, $^{63}$Cu and $^{197}$Au, aiming to get broadly similar statistical uncertainties for all targets.

Based on the simulations performed for E12-20-007 \cite{Li:2020nsk}, and the experience of the earlier $\pi^+$ Hall C nuclear transparency experiment \cite{Clasie:2007aa}, the main physics background within the spectrometer acceptance is expected to come from multi-pion production.  The lower limit for the two pion production phase-space is estimated to be $m_{missing}^2\sim 0.06$ GeV$^2$ for a $^1$H target.  Due to Fermi smearing, the reconstruction resolution will be somewhat worse for the data from heavier nuclei, but this effect can be included in the simulations used to optimize the experimental cuts to be used for each nuclear target.  In \cite{Clasie:2007aa}, the estimated multi-pion background contamination was $<0.4\%$, so it is reasonable to expect this contamination to be no larger than a few percent here.  The contamination of higher mass mesons (such as $\eta$ and $\rho$) should be negligible. The remaining physics background would come from Virtual Compton Scattering (VCS).  While the missing mass reconstruction resolution will not allow the $\pi^0$ and VCS channels to be separated, this contamination is also expected to be $<$1\%.  Thus, the Hall C standard equipment should allow high quality $u$-channel data to be acquired from nuclear targets, and allow nuclear transparency to be studied in backward exclusive $\pi^0$ electroproduction.

\subsection{EIC Perspective}

Despite the difference in the configuration compared to the JLab~12 GeV fixed target experiment, the future Electron-Ion Collider (EIC)~\cite{Accardi:2012qut, EIC:RDHandbook, AbdulKhalek:2021gbh} can be used to probe $u$-channel CT via meson electroproduction: $e+p\rightarrow e^{\prime}+p^{\prime}+\pi^0$ and $e+A(Z)\rightarrow e^{\prime}+p^{\prime} + A^{\prime}(Z-1) +\pi^0$, where $Z$ is the atomic number of the ion beam. To directly extend the kinematics coverage (in $Q^2$) of the JLab measurement, the preferred beam scattering configuration requires a 5 GeV electron beam to collide with a 100 GeV per nucleon ion beam. It is important to note that EIC will offer a variety of ion beams, see details in Ref.~\cite{osti_1765663}. The proposed measurement utilizes the electron and hadron end-caps, and integrated instrumentation in the far forward region (downstream of the outgoing ion beamline). The detection scenario is the following: the scattered electron will be captured by the electron end-cap; the induced virtual photon interacts with a nucleon within the nucleus, then the interacted nucleon transitions into a final state $\pi^0$ through TDA (in $u$-channel kinematics, see Fig.~\ref{Fig_Kinematics_TDAs}); the fast proton (knocked out of the nucleus) will be picked up by the hadron end-cap and create a ``start'' in the timing window; the $\pi^0$ moves out of the nucleus and decays into two photons, thus projecting a one or two-photon signal in the far-forward B0 or Zero Degree Calorimeter. In the case of $eA$ scattering, the $A(Z)$ loses a proton due to the interaction and becomes $A^\prime(Z-1)$, and can be captured by the Roman Pot detector due to the loss of total momentum and magnetic field steering. The feasibility of such a measurement is currently being studied by the EIC ECCE consortium.

Here, it is important to point out that the CT study has been proposed at the EIC through $e+p$ and $e+A$ scatterings, and photoproduction of mesons~\cite{Hauenstein2021,  AbdulKhalek:2021gbh}. These CT studies are based on the validity of collinear factorization theme in the small $-t$ kinematics, and should be distinguished from the $u$-channel meson electroproduction observable proposed in this paper. In the former case, a final state meson will be produced by the $e+p$ and $e+A$ interactions, and will be detected by the central barrel of the EIC; in the latter case, the interacted ion beam and the newly produced meson will both enter the far forward region as described above.

\section{A model estimate of nuclear transparency}

For the $^{12}$C, $^{27}$Al, $^{63}$Cu, and $^{197}$Au nuclei, we provide an estimate for the $A(e,e'p)\pi\, A-1$ nuclear transparency in the backward regime kinematics of Table~\ref{tab:CTkin} available in JLab Hall C.  These estimates are obtained using the Relativistic Multiple Scattering Glauber Approximation (RMSGA).  The RMSGA is a flexible framework that treats kinematics and dynamics (nuclear wave functions, final-state interactions [FSI]) relativistically and has been applied to a variety of hadron-, electron- and neutrino-induced nuclear reactions; see \cite{Ryckebusch:2003fc,Martinez:2005xe,Cosyn:2007er,Cosyn:2013qe} and references therein.  The NT ratio is calculated as 
\begin{equation}
    T = \frac{\sigma^\text{RMSGA}}{\sigma^{PWIA}}\,,
\end{equation}
where in the calculation of the plane wave impulse approximation (PWIA) denominator, FSI are turned off and we integrate nominator and denominator over the experimentally accessible phase space.  The cross-section is calculated in a factorized form~\cite{Cosyn:2007er}:
\begin{multline} \label{eq:cross_A}
    \frac{d\sigma^{eA}}{dE_{e'} d\Omega_{e'} du d\phi_N ds_2 } = \int d\Omega_\pi^* \frac{m_{A-1}}{4s_2}  \sqrt{\frac{\lambda(s_2,m_\pi,m_{A-1})\lambda(s_{\gamma N},m_N,-Q^2)}{\lambda(s_{\gamma A},m_A,-Q^2)}}\\ \times \rho_D(\bm p_i) \; \frac{d\sigma^{eN}}{dE_{e'} d\Omega_{e'} du d\phi_\pi}\,,
\end{multline}
where we integrated over the solid angle of the (undetected) pion in the $(\pi,A-1)$ center-of-mass system.  In the above equation, we have $u=(q-p_N)^2$, and the function
\begin{equation}
    \lambda(s,m_1,m_2) = [s-(m_1-m_2)^2][s-(m_1+m_2)^2]
\end{equation}
was used.
We introduced the invariant masses squared $s_2 = (p_\pi+p_{A-1})^2, s_{\gamma N} = (q+p_i)^2, s_{\gamma A} = (q+p_A)^2$, where $p_N,p_\pi$ and $p_{A-1}$ are the four momenta of the final state proton, pion and remnant $A-1$ nucleus, and
\begin{equation}
    p_i = p_N + p_\pi - q
\end{equation}
is the four momentum of the initial struck nucleon.  The nuclear initial state enters in the distorted momentum distribution
\begin{equation}
    \rho_D(\bm p_i) \equiv \sum_{m_s,\alpha_1} |\bar{u}(\bm p_i, m_s)\phi^D_{\alpha_1}(\bm p_i)|^2\,,
\end{equation}
where the sum over $\alpha_1$ runs over the quantum numbers of the occupied mean-field single particle levels of the initial nucleus $A$.  The wave functions $\phi^D_{\alpha_1}$ include the effects of the FSI between the detected nucleon and the remnant $A-1$ nucleus:
\begin{equation}
    \phi^D_{\alpha_1}(\bm p) = \frac{1}{(2\pi)^{3/2}}\int d^3r \; e^{-i \bm p\cdot \bm r} \;\phi_{\alpha_1}(\bm r) \mathcal{F}^{\text{FSI}}(\bm r)\,.
\end{equation}
 The FSI entering in $\mathcal{F}^{\text{FSI}}(\bm r)$ are parametrized using nucleon-nucleon scattering data, we refer to \cite{Cosyn:2007er} for details.  In the PWIA calculation, we set $\mathcal{F}^{\text{FSI}}(\bm r) \rightarrow 1$.  
The CT effects are implemented through the color diffusion model~\cite{Farrar:1988me,Frankfurt:1994kt}, using $\Delta M^2=1.1 ~\text{GeV}^2$ in the nucleon coherence length $l_h = 2p_N/\Delta M^2$.  The last ingredient of Eq.~(\ref{eq:cross_A}) is the pion production cross-section $\sigma^{eN}$ on the nucleon, which was parameterized in the backward kinematics by interpolating the estimates provided in Ref.~\cite{Li:2020nsk} (see Fig. 19 and Appendix A therein) based on the model of Ref.~\cite{Lansberg:2011aa}.

\begin{figure}
    \centering
    \includegraphics[width=0.8\textwidth]{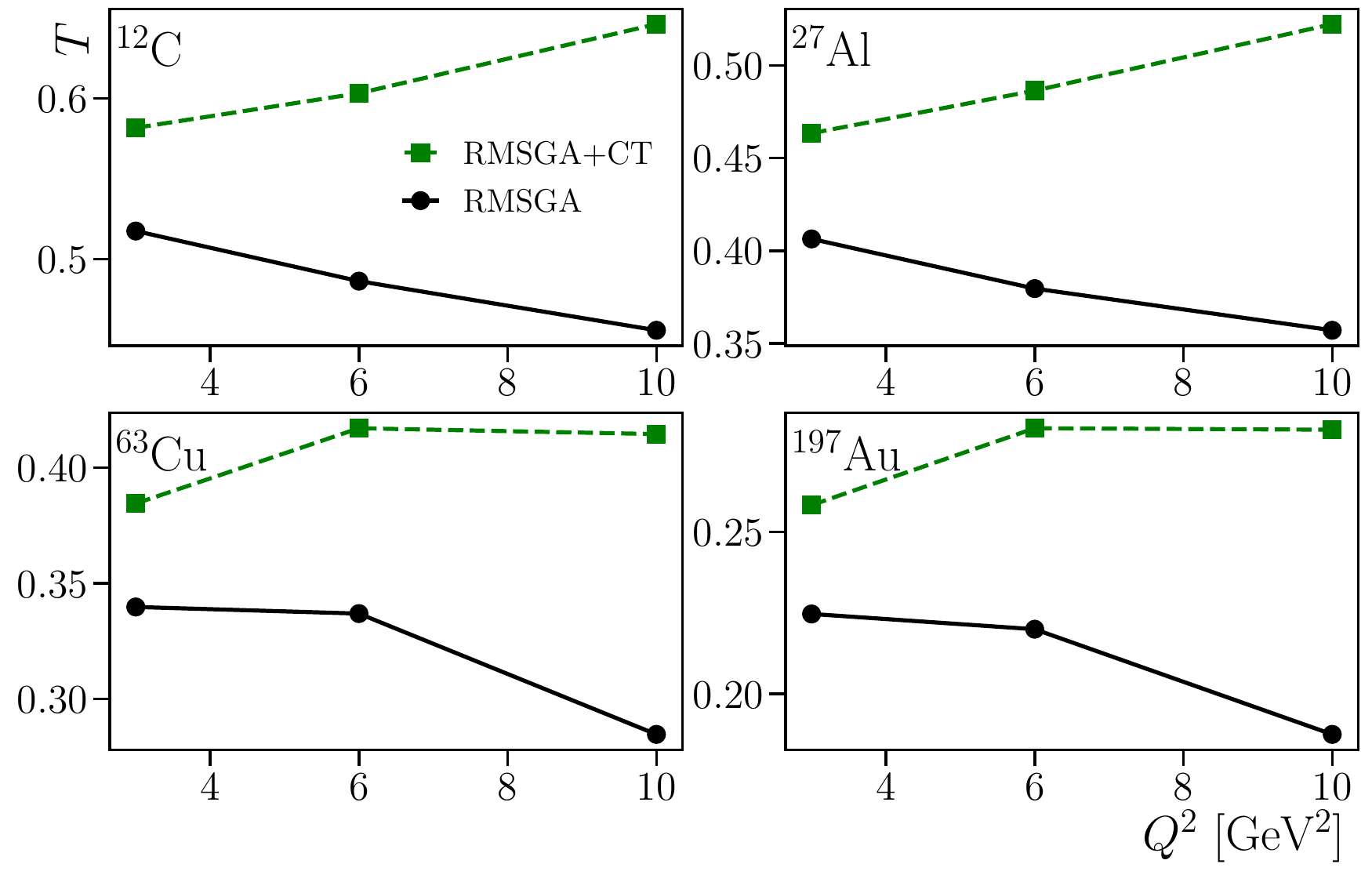}
    \caption{RMSGA nuclear transparency calculations for $^{12}$C, $^{27}$Al, $^{63}$Cu and $^{197}$Au as a function of $Q^2$.  Full curves are regular Glauber RMSGA calculations, the dashed curves include the color transparency in the quantum diffusion model.}
    \label{fig:CT_result}
\end{figure}

Figure \ref{fig:CT_result} shows the results of our calculations, where we took the Table~\ref{tab:CTkin} central values for the final state electron and proton kinematics.  The transparency values lie in the expected range known from $A(e,e'p)$ calculations.  These estimates show that the proposed experiment should be able to distinguish color transparency effects. Note that these predictions can be further improved with a detailed Monte Carlo simulation study and can be extended to EIC kinematics.


\section{Summary}

The available data on nuclear transparency lead to the obvious conclusion that the phenomenon of color transparency needs to be further explored. This is particularly important for the nucleon case. The historical measurement \cite{Carroll:1988rp} of the large angle $pp$ elastic and quasi-elastic$(p,2p)$ scattering have led to many debates and interpretations \cite{Brodsky:1987xw,Ralston:1988rb, Farrar:1988me, Lee:1992rd}. 

The $(e,e^\prime p$) measurements are a natural place to look for proton CT. The recent data on the $(e,e^\prime p)$ reaction~\cite{Bhetuwal:2021} demonstrated the absence of any positive signal for the manifestation of color transparency in this simple reaction up to $Q^2=14$ GeV$^2$, reinforcing doubts on the leading twist dominance of the nucleon form factors at experimentally available energy (especially at JLab 12~GeV). 
With transparency measurements available so far in a limited set of reactions and kinematics, it is too early to tell what drives the absence of the onset of CT in the proton case, while for forward meson production the onset has been observed.
Planned measurements at JLab will extend the known meson production processes to the maximum $Q^2$ values available using the 12 GeV beam at JLab. In this article, we describe a complement to this existing color transparency program, which expands the study to an unexplored territory of $u$-Channel kinematics ($t\rightarrow t_{max}$).

For the first time, backward meson electroproduction on nuclei is linked to CT studies via the collinear QCD factorization framework. It is important to point out that the observation of CT is based on the assumption that a short distance process dominates the amplitude at reachable energies, hence that the TDA predictions are validated by the experimental data.  Observation of CT will help to settle the controversy on the early scaling of exclusive reactions involving nucleons, by answering the question: does the exclusive meson electroproduction experiment witness the dominance of a small nucleon configuration in the backward kinematics where the nucleon has a large energy?

In a quite similar line of thought as in this proposal, one may study nuclear transparency in various experiments (see \cite{Jain:2022xzo} in these proceedings), where TDAs appear as the collinear factorized hadronic matrix element, while a hard scattering process should be accompanied by the appearance of color transparency. This is the case for timelike Compton scattering with a quasi-real photon beam \cite{Pire:2022fbi}, but also in the antiproton nucleus electromagnetic processes at PANDA \cite{Lansberg:2012ha} and the $\pi$-nucleus program at J-PARC \cite{Pire:2016gut}. In these three cases, color transparency should act as a decrease of the initial (rather than final) state interactions.

Once the proposed measurement is completed, a broader discussion within the community is necessary to determine the implication and the global significance of the observed experimental facts, i.e. validation of the TDA formalism, presence or absence of an onset of CT.


%

\section*{Acknowledgments}

We acknowledge numerous fruitful discussions with P. Jain, J.P. Ralston, K. Semenov-Tian-Shansky and L. Szymanowski.  The work of W.C. is partially supported by the National Science Foundation under Award No. 2111442.  The work of G.M.H. is supported by the Natural Sciences and Engineering Research Council of Canada (NSERC), SAPIN-2021-00026.
\bibliography{refs}

%
%
%
%
%
%
%
%

\end{document}